\documentclass[aps,pre,reprint,floatfix]{revtex4-2}
\usepackage{amsmath}
\usepackage{graphicx}
\usepackage{dcolumn}
\usepackage{bm}
\usepackage{siunitx}
\usepackage{color}
\usepackage{soul}

\begin{document}

\title{Hidden dependence of spreading vulnerability on topological
  complexity}

\author{Mark M. Dekker}
\affiliation{Department of Information and Computing Sciences, Utrecht
  University, Princetonplein 5, 3584 CC Utrecht, The Netherlands}
\altaffiliation{Centre for Complex Systems Studies, Utrecht
  University, Minnaertgebouw, Leuvenlaan 4, 3584 CE Utrecht, The
  Netherlands}
\author{Raoul D. Schram}
\affiliation{Information and Technology
  Services, Utrecht University, Utrecht, The Netherlands}
\author{Jiamin Ou}
\affiliation{Department of Sociology, Utrecht University,
  Padualaan 14, 3584 CH, Utrecht, Netherlands}
\author{Debabrata Panja}
\affiliation {Department of Information and Computing Sciences, Utrecht
  University, Princetonplein 5, 3584 CC Utrecht, The Netherlands}
\altaffiliation{Centre for Complex Systems Studies, Utrecht
  University, Minnaertgebouw, Leuvenlaan 4, 3584 CE Utrecht, The
  Netherlands}

\date{\today}

\begin{abstract}
  Many dynamical phenomena in complex systems concern
  spreading that plays out on top of networks with changing
  architecture over time --- commonly known as temporal networks. A
  complex system's proneness to facilitate spreading phenomena, which
  we abbreviate as its `spreading vulnerability', is often surmised to
  be related to the topology of the temporal network featured by the
  system. Yet, cleanly extracting spreading vulnerability of a complex
  system directly from the topological information of the temporal
  network remains a challenge. Here, using data from a diverse set of
  real-world complex systems, we develop the `entropy of temporal
  entanglement' as a novel and insightful quantity to measure
  topological complexities of temporal networks. We show that this
  parameter-free quantity naturally allows for topological comparisons
  across vastly different complex systems. Importantly, by simulating
  three different types of stochastic dynamical processes playing out
  on top of temporal networks, we demonstrate that the entropy of
  temporal entanglement serves as a quantitative embodiment of the
  systems' spreading vulnerability, irrespective of the details of the
  processes. In being able to do so, i.e., in being able to
  quantitatively extract a complex system's proneness to facilitate
  spreading phenomena from topology, this entropic measure opens
  itself for applications in a wide variety of natural, social,
  biological and engineered systems.
\end{abstract}

\maketitle

\section{Introduction}

Networks, consisting of system elements (agents) and their
interactions by nodes and links respectively, have proved to be
effective tools for analyzing complex systems. For a large variety of
them, notable emergent dynamical phenomena of interest concern
spreading, effected by individual agents playing the role of carriers
and transmitting to others as they interact, e.g., sharing information
through conversations \cite{Cattuto2010, PrimarySchool_2}, passing of
signals among animals \cite{Almaas2004, Yosef2011} and of infectious
disease pathogens \cite{Stehle2011b, Firth2020}, purveyance of (fake)
news \cite{Wang2019}, synchronicity in neuronal spikes
\cite{Klausberger2008, Rakshit2018}, and cascading dynamics in
socio-technical systems \cite{Badie2020, Mancastroppa2019, Masuda2020,
  Li2020}. When the agents' identities, and the precise sequences and
timings of their interactions over a given time interval are compiled
together into a temporal network \cite{Holme2012, Riolo2001,
  Isella2011, Perra2012, Li2017, Li2020}, it becomes evident that
spreading is actually a dynamical process taking place on top of the
network.

From this, it seems natural to expect that the topological complexity
of a temporal network will have a profound influence on the system's
proneness to facilitate spreading phenomena --- we abbreviate the
latter as its spreading vulnerability (quantified later in the paper)
--- e.g., the formation and dispersion of epidemiological bubbles or
echo chambers will dictate {\it how fast, or efficiently,\/} pathogens
or (fake) news may spread. However, establishing a quantitative
relation between the two, from which the system's spreading
vulnerability can be extracted directly from the topological
complexity of the temporal network featured by the system, remains a
challenge. The key obstacle stems from the heterogeneities in the
agent interactions intrinsic to temporal networks in the temporal
domain --- there simply are no rules to guide which agent may interact
with which others, when, for how long and in which precise sequence
--- that need to be suitably combined with the topological
complexities that are already present in the network of agent-to-agent
contacts at any given time. Note that the latter, i.e., the agents'
contact network at any given time, is in fact a {\it static\/}
network, for which methods to deal with heterogeneities are already
well-established in the forms of degree distribution, various forms of
centrality measures and community detection algorithms
\cite{Girvan2002}.
\begin{figure*}
    \includegraphics[width=1\linewidth]{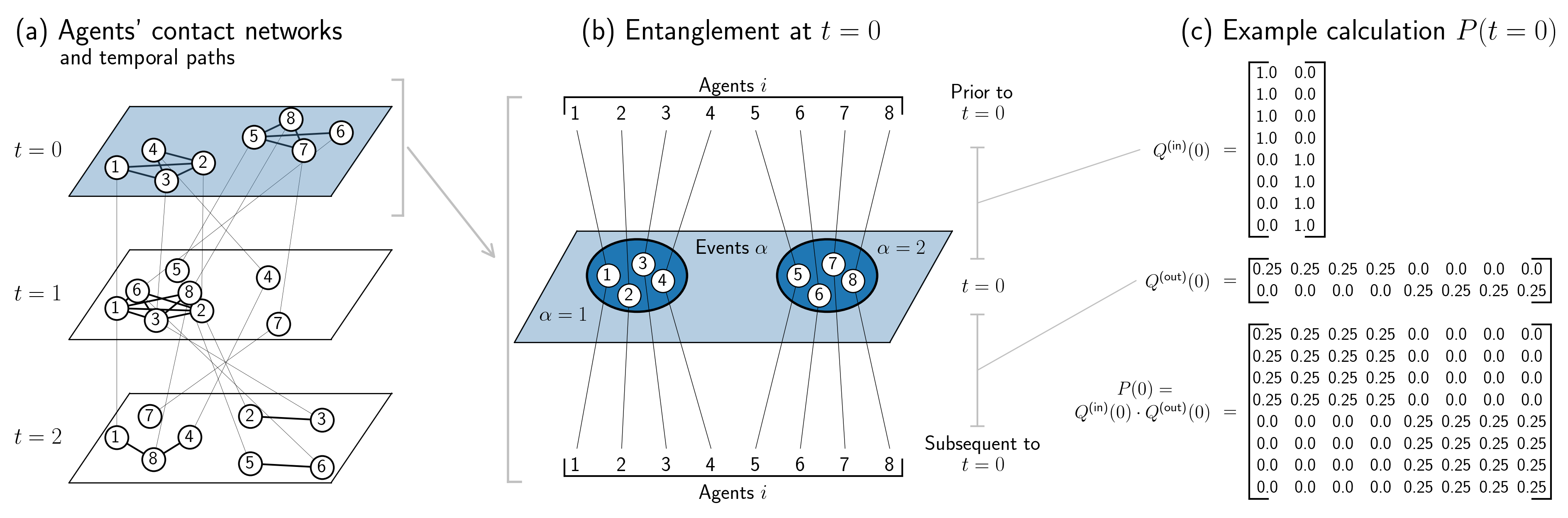}
    \caption{Temporal entanglement.
    \textbf{Panel (a)} An example of a temporal network, in terms of
    agent-to-agent contacts as interactions among eight numbered
    agents, at snapshots taken at integer units of sampling time.
    Snapshots are shown as time layers; the agent-to-agent contacts at
    integer times is shown as black links within the corresponding
    time layer. The agents' temporal paths (strings) from one snapshot
    to the next are shown as gray lines. The changes in the topology
    of the agent-to-agent contacts from one snapshot to the next make
    the strings weave through, and entangle with each other. 
    \textbf{Panel (b)} Entanglement of the strings at $t=0$, extracted
    from panel (a); shown explicitly are the temporal paths (strings),
    labeled by Roman agent indices, and the events in dark blue
    circles, labeled by Greek indices. Events are the connected
    components of the agents' contact network at $t=0$. \textbf{Panel
      (c)} Calculation of the agent-to-agent propagator matrix $P(0)$
    as $P(0)=Q^{(\text{in})}(0)\cdot Q^{(\text{out})}(0)$. The
    $i\alpha$-th element of the agent-to-event propagator matrix
    $Q^{(\text{in})}(0)$ is the probability of a random walker to
    start at agent $i$ prior to $t=0$ and end at event $\alpha$ at
    $t=0$ in one hop. Similarly, the $\alpha i$-th element of the
    event-to-agent propagator matrix $Q^{(\text{out})}(0)$ is the
    probability of a random walker to start at event $\alpha$ at $t=0$
    and end at agent $i$ subsequent to $t=0$ in one hop.}
    \label{fig1}
\end{figure*}

Here we develop an insightful quantity to measure topological
complexities of temporal networks: we name it the \textit{entropy of
  temporal entanglement} (the choice for this name is justified below,
and it bears no connection to entanglement entropy for quantum
many-body systems). Using real-world temporal networks data, and
simulating three different dynamical processes on top of them, we show
that this entropic measure not only allows for a quantitative
comparison of vastly different complex systems, but crucially, it also
bears a clean relation to the system's spreading vulnerability,
irrespective of the details of the processes. In other words, this
entropic measure is able to directly, and cleanly, extract a complex
system's proneness to facilitate spreading phenomena from its
topology.

We coin the term `entanglement' because of the following. Real-world
temporal networks often follow a `discrete time convention', wherein
the agents' interactions, denoted by agent-to-agent contacts, are
sampled at some fixed interval (which we denote as $\tau_s$): an
example is shown in Fig.~\ref{fig1}(a), where the temporal network
consists of layers in time, denoted in integer units of $\tau_s$. Then
the gray lines denoting the agents' temporal paths across the layers
in Fig.~\ref{fig1}(a) --- obtained by following individual agents in
time --- resemble strings laid out in time. Changes in agent
interactions, i.e., in agent-to-agent contacts, from one snapshot to
the next make the strings weave through and {\it entangle\/} with each
other. (In the absence of any interactions among the agents at integer
times, the strings will simply be parallel to each other, without any
entanglement.)

\section{Summary calculation of the entropy of temporal entanglement \label{sec2}}

We begin by introducing `events'. Events are the connected components
of the agents' contact networks at integer times. An example can be
found in Fig.~\ref{fig1}(a-b), where the interactions among agents 1
through 4 and agents 5 through 8 constitute two distinct events at
$t=0$. In this definition, every agent becomes a part of a single
event, and all agents are equivalent within an event; in addition,
some may be solo-agent events, as seen at $t=1$ in Fig.~\ref{fig1}(a).
[Note also that the equivalence of agents within an event does not
respect the precise agent-to-agent contacts within events, such as the
absence of a direct contact between agents 1 and 4 at $t=0$ in
Fig.~\ref{fig1}(a), while all the others are in direct contact in that
event; we take this up in the paragraph below Eq. (\ref{eq:S}).]

We then probe the topology of a temporal network, consisting of $N$
agents, using random walkers hopping along the agents' temporal paths
forward in time. (Probing network topology using random walkers is a
common practice.) The entire procedure is parameter-free, and is
formally described in SI A.1. We summarize it here for $m$ events at
time $t$ in three steps; a rendering of them for the example system in
Fig.~\ref{fig1}(a) with $N=8$ and $m=2$ at $t=0$, and the
corresponding calculations are shown in Fig.~\ref{fig1}(b-c). (1)
Using Roman and Greek letter indices to denote agents and events
respectively, we construct the $N\times m$ `agent-to-event propagator
matrix' $Q^{(\text{in})}(t)$, with matrix element
$Q^{(\text{in})}_{i\alpha}$ denoting the probability for a random
walker to start at agent $i$ prior to $t$ and reach event $\alpha$ at
(integer) time $t$ in one hop. (2) Similarly, we construct the
$m\times N$ `event-to-agent propagator matrix' $Q^{(\text{out})}(t)$,
with matrix element $Q^{(\text{out})}_{\alpha i}$ denoting the
probability for a random walker, to start at event $\alpha$ at time
$t$ and reach agent $i$ subsequent to time $t$ in one hop. (For both
steps, the hops are thus coupled to the sampling time interval
$\tau_s$.) (3) Finally, we construct the $N\times N$ `agent-to-agent
propagator matrix' as
\begin{eqnarray} P(t) &=&
                          Q^{(\text{in})}(t)\cdot Q^{(\text{out})}(t).
\label{e1}
\end{eqnarray}

Upon extending this procedure to finite (integer) interval $\Delta t$
leads us to the product matrix $\wp(t, \Delta t)=P(t)\cdot P(t+1)\cdot
P(t+2)\ldots P(t+\Delta t)$ that similarly contains the full
entanglement information in the finite time interval $[t,t+\Delta t]$.
By construction, the matrix element $\wp_{ij}(t,\Delta t)$ is the
probability of a random walker starting at agent $i$ prior to time $t$
to end up at agent $j$ subsequent to time $(t+\Delta t)$ following
the agents' temporal paths. This allows us to define 
\begin{eqnarray}
\label{eq:sj}
s_i(t, \Delta t) &=& -\sum_j \wp_{ij}(t, \Delta t)\,
                   \ln\, \wp_{ij}(t, \Delta t)
\end{eqnarray}
as the (information-theoretic) entropy assosiated with the random
walker starting at agent $i$ prior to time $t$. A further sum over
$i$, weighing each starting random walker equally, provides a
network-wide (global) sum of the $s_i(t,\Delta t)$ entropies, allowing
us to define the entropy of temporal entanglement for all agents over
the interval $[t,t+\Delta t]$ as
\begin{eqnarray}
\label{eq:S}
S(t, \Delta t) &=& -\frac{1}{N\ln N}\sum_{i,j} \wp_{ij}(t, \Delta t)\,
                   \ln\, \wp_{ij}(t, \Delta t). 
\end{eqnarray}
Given that $s_j(t, \Delta t) $ is upper bounded by $\ln N$ and the
equal weight given to each of the $N$ random walkers for calculating
the entropy of temporal entanglement, $N\ln N$ factor in the
denominator of $S(t, \Delta t)$ ensures $S(t, \Delta t)\in[0,1]$.

Construction of the entropy of temporal entanglement that respects the
precise agent-to-agent contacts within events [such as the absence of
a direct contact between agents 1 and 4 at $t= 0$ in
Fig.~\ref{fig1}(a), while all the others are in direct contact in that
event] follows a line similar to the above, albeit it is slightly more
involved. The corresponding entropy of temporal entanglement
$S_{\text c}$, formally derived in SI A.2, is also parameter-free and
fully determined by the topology of the network. In SI A.2 we argue
that with decreasing sampling time interval $\tau_s$, $S_{\text c}$
approaches $S$, the entropy of temporal entanglement as calculated in
Eqs. (\ref{e1}-\ref{eq:S}) in terms of the connected components of the
agents' contact networks, which we henceforth adhere to.

\section{Properties and interpretation of the entropy of temporal entanglement\label{sec3}}

\begin{figure*}
    \includegraphics[width=1\linewidth]{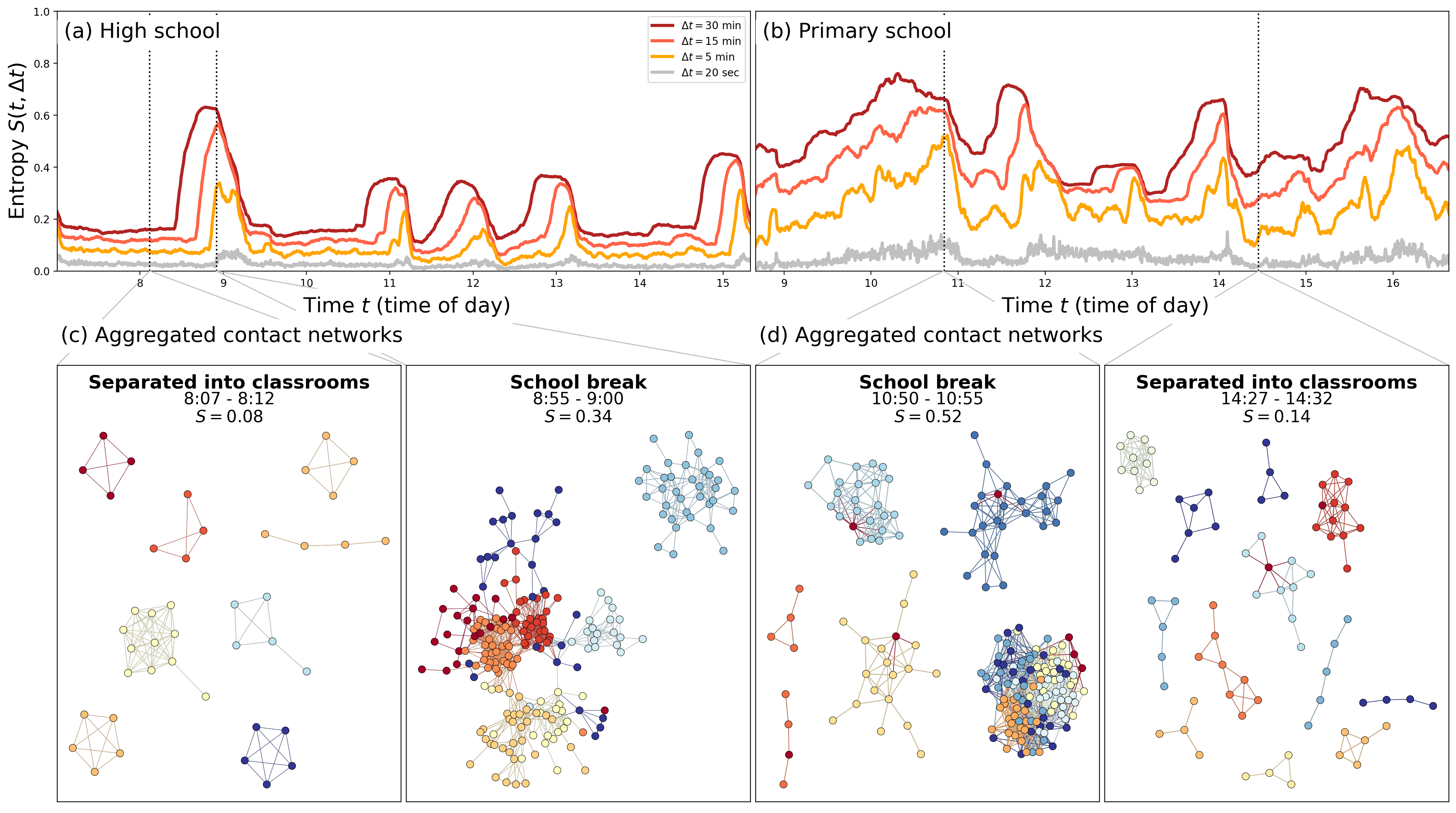}
    \caption{Entropy of temporal entanglement for two real-world
      temporal networks: contacts among high school (Thiers13 dataset
      day 2, left column) and primary school (LyonSchool dataset day
      1, right column) students. \textbf{Panels (a) and (b)}
      $S(t, \Delta t)$ for four values of $\Delta t$: 20 seconds
      (gray), 5 minutes (orange), 15 minutes (red) and 30 minutes
      (darker red) --- for any value of $t$, note that
      $S(t, \Delta t)$ is monotonically increasing with $\Delta t$ ---
      $x$-axes denote the time of day. Vertical dashed lines
      correspond to 5-minute intervals, expanded in panels (c) and
      (d), chosen to reflect one high-entropy and one low-entropy
      situation per panel. \textbf{Panels (c) and (d)} Aggregated
      contact networks within the 5-minute intervals (only connected
      components with at least 4 students are shown), demonstrating
      high (resp. low) connectivity in high-$S$ (resp. low-$S$) cases.
      Metadata reveal that the situations correspond to school breaks
      and ongoing classes respectively (students belonging to the same
      classroom are represented by the same node color).}
    \label{fig2}
\end{figure*}

Three insightful attributes of $S(t,\Delta t)$ are critical for being
able to extract a complex system's proneness to facilitate spreading
phenomena. These are: (1) $S(t,\Delta t)$ represents the agents'
mixing (i.e., mingling) propensity in the interval $[t,t+\Delta t]$
(we explore this further in \ref{sec3_2} below), (2) it respects the
{\it temporal sequence\/} of agent interactions, and (3) in building
up the sequences forward in time, it automatically incorporates {\it
  causality\/}: note that a spreading phenomenon in the past can
influence its future dynamics, but not {\it vice versa\/}. While (2-3)
can be gleaned from how $\wp$ is built from the $P$-matrices, (1)
follows from that $\wp_{ij}(t,\Delta t)$ is the probability of a
random walker starting at agent $i$ prior to time $t$ to end up at
agent $j$ subsequent to time $(t+\Delta t)$ following temporal paths:
the more distinct temporal paths there are to trace a given agent $i$
back from any agent $j$, the more nonzero $\wp_{ij}$ elements there
are. These attributes further translate to the following useful
properties of $S(t,\Delta t)$. (a) An agent, who never interacts with
another in the interval $[t,t+\Delta t]$, has zero contribution to
$S(t,\Delta t)$. This, in fact, is the basis for why small
$S(t, \Delta t)$ corresponds to a highly fragmented temporal network,
i.e., large number of disjoint components in the interval
$[t,t+\Delta t]$, possibly with many agents separated from each other
throughout the interval. (b) It allows us to quantify the contribution
of an individual agent to entanglement, which we take up in SI~A.3.
(c) Typically, an increasing $\Delta t$ would imply more nonzero
$\wp_{ij}$'s: more distinct temporal paths become available to trace a
given agent $i$ back from any agent $j$. This implies that for any
value of $t$, $S(t, \Delta t)$ is monotonic in $\Delta t$; e.g., as
can be seen in Fig.~\ref{fig2}(a-b); for a fixed value of $\Delta t$
however, $S(t, \Delta t)$ is non-monotonic in
$t$.

In contrast to the above, the practice of aggregating the agents'
contact networks over a certain time interval for a temporal network
into a static one \cite{Li2020, Isella2011, Riolo2001, Masuda2013,
  PrimarySchool_2} discards the precise sequences and durations of
agent interactions. As for methods that do capture heterogeneities in
the temporal domain by means of, e.g., checking how communities evolve
in time \cite{Rosvall2010, Peixoto2017}, constructing temporal
counterparts of degree and betweenness \cite{Kim2012}, or treating a
temporal network as a multilayer one with layers representing discrete
time snapshots \cite{Mucha2010}, we note the following. The emphasis
of temporal betweenness on shortest paths makes it ideal to identify
topological centrality, but for the application of spreading
vulnerability, path \textit{lengths} are only important in terms of
which agents are traceable in a fixed time interval. Of critical
importance for spreading vulnerability are the exact interaction
sequences and frequencies, which are considerably less focused on in
the existing literature. Likewise, concerning community detection
approaches, under time reversal, the community structure in the
multilayer network will remain invariant, while the dynamics of a
process taking place on top of the temporal network will be profoundly
different.

\subsection{Entropy of temporal entanglement represents the agents' mixing propensity\label{sec3_1}}

In order to demonstrate that the entropy of temporal entanglement has
a clear relationship to the agents' propensity to mixing, we consider
two temporal network datasets that track student contacts respectively
in a high school and a primary school by wearable sensors, sampled at
20 seconds intervals \cite{Genois2018, PrimarySchool_2, HighSchool_1}
(details on the data in SI C). Per dataset, in Fig.~\ref{fig2}(a-b) we
plot the time evolution of $S(t, \Delta t)$ for four different
$\Delta t$-values, varying $t$ over one full school day. Further,
delving into the (publicly available) metadata, in
Fig.~\ref{fig2}(c-d) we plot the contact networks of the agents
aggregated over two five-minute intervals, one each for high and low
values of $S(t, \Delta t)$. The correspondence between the top and the
bottom panels demonstrates that the students' mixing behavior,
adapting to changing school circumstances, is being reflected in the
topological complexity of the temporal network, which in turn is being
captured by the entropy of temporal entanglement (school breaks invite
more mixing in comparison to being separated in classrooms, and
correspondingly lead to more topological complexity and higher entropy
of temporal entanglement).
\begin{figure*}
    \includegraphics[width=1\linewidth]{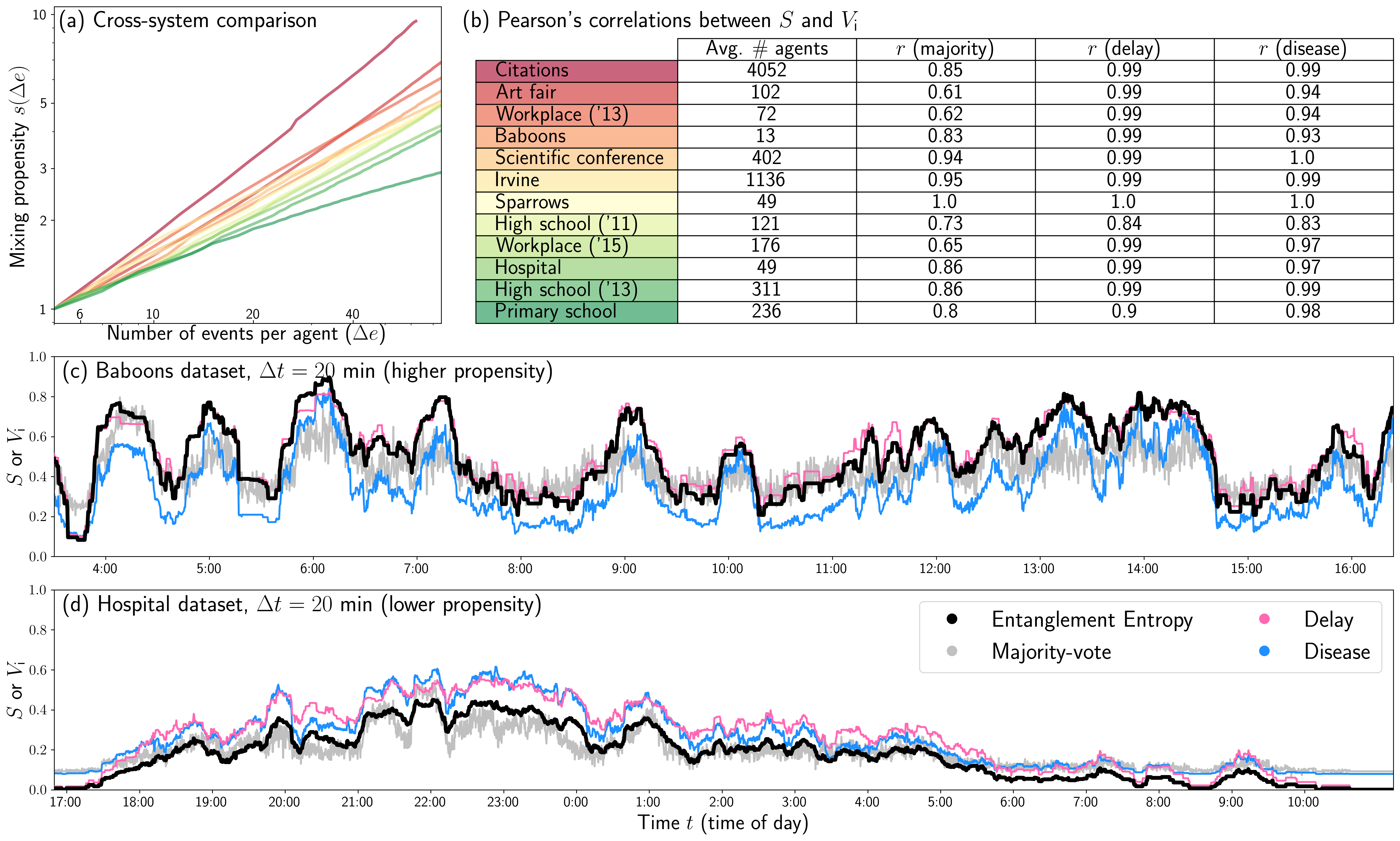}
    \caption{Standardized mixing propensity $s$ and the systems'
      vulnerability $V$ to spreading phenomena (for precise
      definitions of both, see main text \ref{sec3_2} below).  
    \textbf{Panel (a)} Mixing propensity vs the number of events per
    agent as a system-characterizing property, shown for 12 complex
    systems: from green (slow rate of increase) to red (fast rate of
    increase). The systems are noted in panel (b) in the same color
    coding. 
    \textbf{Panel (b)} Pearson's correlation between $S(t,\Delta t)$
    and $V_{\text{i}}(t, \Delta t)$ , with
    $\text{i}\in(\text{maj},\text{del},\text{dis})$, corresponding to
    a majority-vote model, a transport delay model and an
    epidemiological susceptible-infected (SI) model respectively (see
    text and SI C for details). 
    \textbf{Panel (c)} The full time-series of $S(t,\Delta t)$ and $V_{\text{i}}(t, \Delta t)$
    on day 2 of the Baboons dataset \cite{baboons}. Displayed on the
    $x$-axis are the times of day. $\Delta t = 20$ minutes (i.e., 60
    units of sampling time intervals) is used. 
    \textbf{Panel (d).} Same as in panel (c), but using day 2 of the
    Hospital dataset \cite{hospital}.}
    \label{fig3}
\end{figure*}


Given the above, an unbiased (e.g., from number of agents, sampling
time interval) cross-system comparison of topological complexities of
temporal networks requires two rescaling operations. The first
(natural) one is that $\Delta t$ must be replaced by $\Delta e$, the
average number of events per agent in the interval $\Delta t$ (the
assumption here is that the agents' contact networks are sampled
frequently enough such that contact sequences are fully captured). The
second one concerns normalizing the entropy to define a standardized
mixing propensity: 
\begin{eqnarray}
\label{eq:prop}
s(\Delta e) &=& \frac{1}{T}\sum_{t=0}^T\frac{S(t, \Delta e)}{S(t, \Delta e_0)},
\end{eqnarray}
where $\Delta e_0$ is a constant for standardization. Note that
$s(\Delta e)$ is a monotonically increasing function of $\Delta e$
since $S(t, \Delta t)$ is a monotonically increasing function of
$\Delta t$. Plotting $s(\Delta e)$ vs. $\Delta e$ in a log-log plot in
Fig.~\ref{fig3}(a), using $\Delta e_0=5$, we observe the differences
in how characteristically fast $s(\Delta e)$ builds up in different
real-world systems (see SI C for details on the systems). (Even though
$\Delta e_0$ is merely a standardization parameter, it should be
chosen not too small to have some stability in the plot. We have
experimented with other values such as $\Delta e_0 = 1$ and the
results do not substantially differ.) We analyze this further in SI B
to demonstrate that the fastest rate of increase in the entropy of
temporal entanglement is achieved when the agents' contact sequences,
in the temporal domain, only contain trees. Real-world temporal
networks of course contain loops (e.g., due to repetitive interactions
among agents); note also that social contacts are often structured in
``bubbles'', which enhances the chances of having loops.

\subsection{Entropy of temporal entanglement is an embodiment of systems' vulnerability to spreading phenomena\label{sec3_2}}

We begin by quantifying spreading vulnerability of a complex system,
which requires adding dynamical processes on top of the temporal
network that the system features. For this purpose, we consider three
different types of stochastic processes: (1) a majority-vote model
that simulates the opinion on a dilemma that spreads through events by
means of the majority votes, (2) a transport delay model, where delay
spreads through events due to all the participants copying the delay
of the maximally delayed agent, and (3) an epidemiological
susceptible-infected (SI) model with infection probability
$\beta=0.8$. Details of the models and their backgrounds can be found
in SI D. The corresponding process variables are respectively
expressed as $V_{\text{i}}(t, \Delta t)$, with
$\text{i}\in(\text{maj},\text{del},\text{dis})$, and just like the
entropy of temporal entanglement $S(t,\Delta t)$, they are normalized
to the interval $[0,1]$ (elaborated in SI D). Like $S(t,\Delta t)$,
for any given $t$, $V_{\text{i}}(t, \Delta t)$ is a monotonically
increasing function of $\Delta t$, and the spreading vulnerability is
quantified by how fast $V_{\text{i}}(t, \Delta t)$ increases as a
function of $\Delta t$. [Also, like $S(t,\Delta t)$,
$V_{\text{i}}(t, \Delta t)$ exhibits strong heterogeneous behavior as
a function of $t$.]

We then simulate the above three stochastic processes for all the
systems in Fig.~\ref{fig3}(a). Using $\Delta t=15$ minutes we plot the
process variables together in Figs.~\ref{fig3}(c-d) for the temporal
network dataset for interacting baboons \cite{baboons} and hospital
patients and workers\cite{hospital}: visual inspection immediately
reveals that for both systems, the variations in $S$ and
$V_{\text{i}}$ over time are highly synchronized. For all the complex
systems in Fig.~\ref{fig3}(a), the Pearson's correlation coefficients
between $S(t,\Delta t)$ and $V_{\text{i}}(t, \Delta t)$ are tabulated
in Fig.~\ref{fig3}(b) using the same color scheme [of
Fig.~\ref{fig3}(a)]. The table clearly showcases that the entropy of
temporal entanglement is essentially an embodiment of a complex
system's vulnerability to spreading phenomena, irrespective of the
details of the processes playing out on top of the networks. In other
words, the entropy of temporal entanglement secures a clean relation
between the spreading vulnerability of a complex system, and the
topology of the temporal network featured by the system, and thereby,
the current work indicates that computing the entropy of temporal
entanglement is sufficient to quantitatively assess a complex system's
proneness to facilitate spreading phenomena.


\section{Discussion and outlook\label{sec4}}


Summarizing, we have developed the entropy of temporal entanglement as
a measure of topological complexity of temporal networks. By
construction, it is a parameter-free quantity that embodies collective
topological property at any given timescale, allowing for comparisons
across vastly different complex systems. By simulating three different
dynamical processes playing out on the top of real-world temporal
networks, we have demonstrated that the entropy of temporal
entanglement is a good representation of the systems' spreading
vulnerability, irrespective of the details of the processes --- less
entanglement (lower topological complexity) means lower spreading
vulnerability and {\it vice versa\/}.

Let us reflect on why this topological entropy measure is such a good
predictor of spreading dynamics, irrespective of the details of the
dynamical processes put on top on the temporal networks. The reason
can be traced back to the fact that the $\wp_{ij}(t,\Delta t)$ is the
probability of a random walker to start at agent $i$ prior to time
$t$, and end up at agent $j$ subsequent to time $t+\Delta t$. With
$\sum_j\wp_{ij}(t,\Delta t)=1$, the time evolution of
$\wp_{ij}(t,\Delta t)$ as a function of $\Delta t$ can itself be seen
as a spreading process starting at agent $i$ at time $t$, albeit a
{\it deterministic\/} one --- determined entirely by the topology of
the temporal network, free of any parameter. It is therefore logical
that the entropy of temporal entanglement serves as a quantitative
embodiment of the systems' spreading vulnerability, irrespective of
the details of the processes. Nevertheless, the important fact remains
that the entropy of temporal entanglement establishes a clean
quantitative relation between the topological complexity of a temporal
network and the system's proneness to facilitate spreading phenomena.

\begin{figure}[ht]
  \begin{center}
    \includegraphics[width=0.45\linewidth]{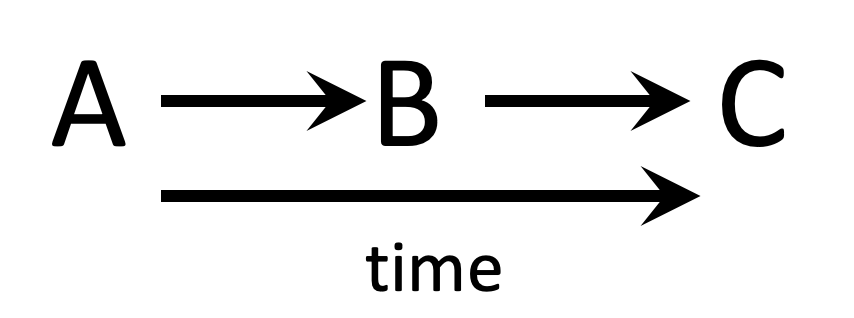}\hspace{8mm}\includegraphics[width=0.45\linewidth]{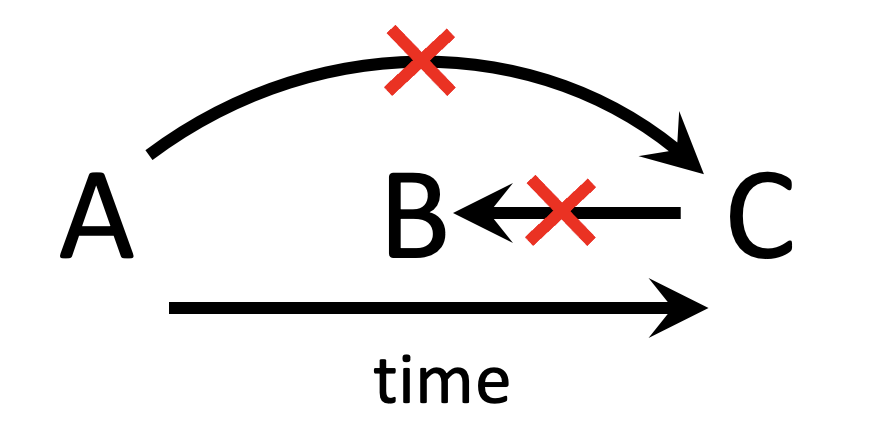}
  \end{center}
 \caption{Three agents A, B and C interacting in
       sequence. (Left) A propagation process that is in reality
       allowed, (Right) a propagation process that is in reality
       impossible, but is allowed when the agent interactions information
       in the temporal dimension is collapsed.\label{fig4}}
\end{figure}

That said, based on Fig. \ref{fig3}(b-d), one may potentially argue
along the following lines. Instead of a sophisticated quantity like
the entropy of temporal entanglement, why not simply use
$S'(t, \Delta t) = (\text{number of contacts between}\, t\,
\text{and}\, \Delta t)/[N(N-1)/2]$ to demonstrate that the speed of
spreading in the interval $[t,t+\Delta t]$ is a direct reflection of
the temporal density of agent-to-agent contacts? This is a subtle
issue, and can be addressed as follows. In calculating $S'$, all the
information on the agent interactions in the temporal dimension
between $t$ and $t+\Delta t$ is collapsed, which erases all
information on the {\it sequence\/} of agent interactions (we have
mentioned this briefly in the paragraph above Sec. \ref{sec3_1}). For
a spreading process, the sequence is critical: if agent A interacts
with agent B, followed by B interacting with C, then the spreading
process must follow the sequence A $\rightarrow$ B $\rightarrow$ C,
while in the network obtained by collapsing the temporal dimension,
since the sequential information of the agent interactions is lost,
spreading can also take place from A $\rightarrow$ C $\rightarrow$ B
(Fig. \ref{fig4}).

The key point this paper posits is that for a temporal network, the
topological counterpart of the spreading process respecting the
interaction sequence is the entropy of temporal entanglement. It is
for this reason that $S(t,\Delta t)$, in comparison to
$S'(t,\Delta t)$, is not simply capturing the higher-order structures
in a temporal network, but is actually respecting the correct
sequential information of agent interactions in the topology of the
network. This issue have recently been discussed more extensively in a
recent paper co-authored by three of us \cite{borsboom}.


System vulnerability is receiving progressively more attention in
widely diverse areas, especially in the context of large perturbations
induced by, e.g., the COVID-19 pandemic, species' habitat loss and
fragmentation, disruption of food webs, and robustness of biological
systems such as gene regulation, metabolism, neural dynamics, and
engineered systems. The common features underlying this diversity are
that (a) these are complex systems, wherein the components (e.g.,
species, genes) have (developed) time-varying functional dependencies
among each other (e.g., sharing resources and by-products, regulating
biochemistry) that can be expressed as temporal networks, and (b) the
effect of a perturbation, albeit initially localized in space and
time, will potentially spread or cascade through the entire system.
Predicting the spreading vulnerability of perturbations directly from
topological complexity would be an asset --- especially considering
that the availability of real-world data on large perturbations are
typically limited --- for example, for mapping their tipping points
\cite{dakos2019}, or alternatively, for tracing the boundaries of
their safe operating spaces \cite{rockstrom2009} that separate
successful recovery from irreversible degradation. 

Finally, we note that all the processes we considered in the paper
for spreading dynamics are so-called ``point processes'', meaning that
they are processes that do not contain internal (memory-type)
time-scales. In real world however, dynamical processes running on top
of the (temporal) networks will undoubtedly possess a diverse range of
time-scales (e.g., incubation period or time to become infectious for
individual agents in case of pathogen spreading), and the process
time-scales will interact with the time-scales of the changes in the
network topology. We expect that the topological complexity of the
networks to still be a crucial factor in dictating the spreading
dynamics: the general nature of our methodological approach sets the
stage for these applications in such wide and far-reaching areas, some
of which are in the process of currently being investigated.


\section*{Acknowledgement}

The authors thank Denny Borsboom, Michael X Cohen, Sander van Doorn,
Hans Heesterbeek, Mirjam Kretzschmar and Paul van der Schoot for their
useful remarks on the manuscript. The work has been financially
supported by Dutch Research Council (NWO), and co-supported by
Nederlandse Spoorwegen (NS) and ProRail, under project number
439.16.111.

All authors contributed to the research conceptualization. MMD, DP and
RS conceived the mathematical principles. MMD did the data analysis,
with help of JO. MMD wrote the code for the data analyses and
simulations, with help of RS. MMD and DP wrote first draft of the
manuscript. All authors reviewed the final text.

\section*{Appendix}

The Appendix consist of four sections. \ref{sec_a1} contains the
formal derivation of the entropy of temporal entanglement.
\ref{sec_a2} contains the behavior of the entropy of temporal
entanglement as a function of $\Delta t$. \ref{sec_a3} contains
information on the real-world temporal network datasets used in the
paper. In \ref{sec_a4} we elaborate on spreading processes playing out
on top of temporal networks.

Detailed methods and information on the accessibility of the data can
be found in the Supplementary Information files. In particular, the
data needed to perform the analyses in the paper are all open source
and discussed in \ref{sec_a3} .

\renewcommand{\thesection}{Appendix \Roman{section}}
\renewcommand{\theequation}{A\arabic{equation}}
\setcounter{section}{0}
\counterwithin{figure}{section}

\section{Formal derivation of the entanglement entropy\label{sec_a1}}

In this Section, we derive the building blocks $Q^{\text{(in)}}$ and
$Q^{\text{(out)}}$ matrices for calculating the entanglement
entropy. In \ref{sec_a1}, we do this for connected components of the
network. In \ref{sec_a2}, we follow this up with the derivation of
$Q^{\text{(in)}}$ and $Q^{\text{(out)}}$ matrices for the agents'
precise contact networks at the snapshots as recorded in the data
[e.g., as can be seen in Fig.~1(a) at $t=0$ in the main text], which
we henceforth refer to as ``fine structures'' for brevity. We will
witness that the $Q$-matrices for both cases are determined from
topological symmetry considerations, and are free of fitting
parameters. The contribution of individual agents to entanglement is
defined in A.3.

\subsection{Entanglement entropy for connected components}

\begin{figure*}
  \begin{center}
      \includegraphics[width=1\linewidth]{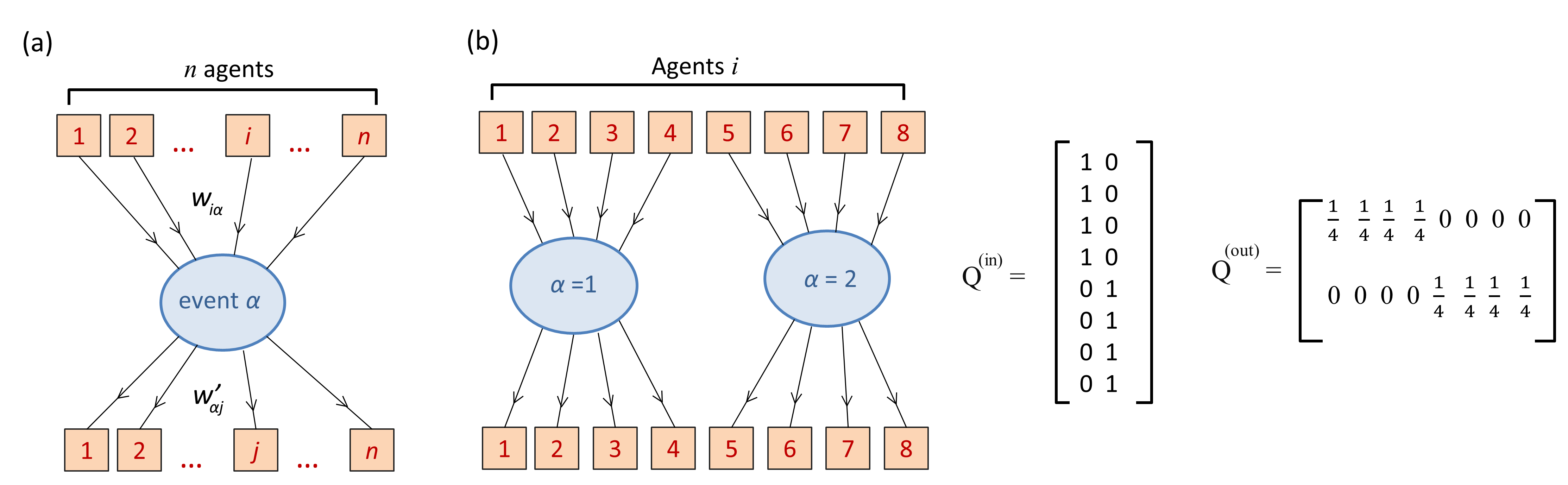}
  \end{center}
  \caption{(a) Event $\alpha$ taking place at time $t$ for one of the
    connected components with $n$ agents, labeled by Roman
    indices. (b) As a working example we consider the same event
    structure for eight agents at $t=0$ as in Fig. 1(b) in the main
    text of the paper. \label{fragment}}
\end{figure*}

Let us start with a single connected component at time $t$ [for
example as shown in Fig. 1(b) in the main text]. For this, it means
that there is a single event $\alpha$ containing $n$ agents at time
$t$. Following our convention regarding the agent strings as explained
in the main text surrounding Fig. 1, such a situation is shown in
Fig.~\ref{fragment}(a). Therein, we denote the (topological) weight of
a link from agent $i$ coming into event $\alpha$ by $w_{i\alpha}$, and
similarly, the topological weight of a link exiting event $\alpha$ and
reaching agent $j$ by $w'_{\alpha j}$. These weights are the elements
of the $Q^{\text{(in)}}$ and $Q^{\text{(out)}}$ matrices,
respectively. In this section, we determine these weights from the
following considerations, and subsequently construct the corresponding
$Q$ and $P$ matrices.

\begin{itemize}
\setlength\itemsep{-1.5mm}
\item[A1.] The first observation is that the entanglement topology is
  invariant under time reversal. This implies
\begin{eqnarray}
\frac{w_{1\alpha}}{w'_{\alpha1}}=\frac{w_{2\alpha}}{w'_{\alpha2}}=\ldots=\frac{w_{n\alpha}}{w'_{\alpha n}}=c_\alpha,
\label{eA1}
\end{eqnarray}
for some (yet unknown) $c_\alpha$.

\item[A2.] The second observation is that the entanglement topology is
  invariant under an exchange operation $i\leftrightarrow j$ for all
  pairs of the agent identities $i$ and $j$. This further implies
  $w_{1\alpha}=w_{2\alpha}=\ldots=w_{n\alpha}=w$, for some (yet
  unknown) $w$.
  
\item[A3.] Finally, $w$ and $c$ are determined by considering random
  walker hops, as is standard in Network Science, in the following
  manner. (a) First, $w_{i\alpha}$ is the probability for a random
  walker to start at agent $i$ prior to time $t$ and end at event
  $\alpha$ at time $t$ in one hop, meaning that $w=1$. (b) Similarly,
  $w_{\alpha j}$ is the probability of a random walker to start at
  event $\alpha$ at time $t$ and end at agent $j$ subsequent to time
  $t$, yielding $c_\alpha=n$. Returning to the original notation from
  Fig. \ref{fragment}(a), this means that $w_{i\alpha}=1\,\,\forall i$
  and $w'_{\alpha j}=1/n\,\,\forall j$.
\end{itemize}
Important to note here is that the random walkers only move forward in
time; i.e., the link weights respect the direction of time.
\begin{figure}[h]
  \begin{center}
    \includegraphics[width=0.8\linewidth]{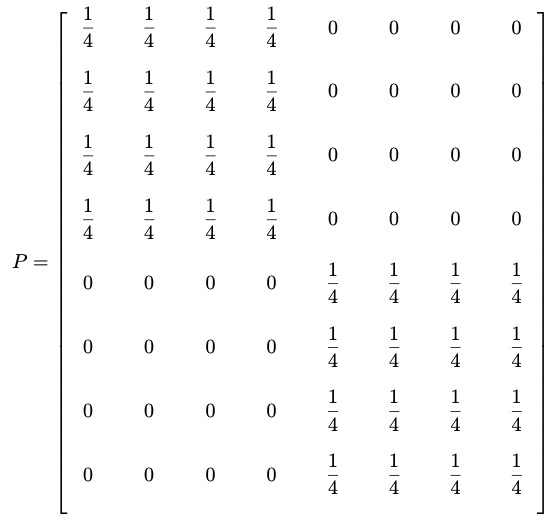}
    \caption{The $8\times8$ $P$-matrix constructed as
      $P=Q^{\text{(in)}}\cdot Q^{\text{(out)}}$ from
      Fig.~\ref{fragment}(b).\label{pfmatrix}}
\end{center}
\end{figure}

Upon putting A1-A3 together, we can construct the relevant
$Q^{(\text{in})}, Q^{(\text{out})}$ and $P$ matrices for this
connected component. First, $Q^{\text{(in)}}$, whose $i\alpha$-th
element equals $w_{i\alpha}$, becomes an $n\times1$ matrix with all
entries unity, and $Q^{\text{(out)}}$, whose $\alpha j$-th element
equals $w'_{\alpha j}$, becomes a $1\times n$ matrix with all elements
equaling $1/n$.  Next, $P=Q^{\text{(in)}}\cdot Q^{\text{(out)}}$
becomes an $n\times n$ matrix, for which all elements are equal to
$1/n$, meaning that at time $t$ all involved ($=n$) agents contribute
$n\ln n$ to the entanglement entropy. [Note here that both
$Q^{\text{(in)}}$ and $Q^{\text{(out)}}$ are row-normalized, which
makes $P$ both row- and column-normalized. These normalizations stem
from the conservation of random walkers --- any random walker starting
from any agent prior to time $t$ must end up at some agent subsequent
to time $t$.] In other words, the agents participating in the event
are {\it maximally entangled\/} at time $t$: indeed, if at time
$(t+1)$ the same $n$ agents also participate in an event together that
involves no new agents, then ensured by the relation $P^2=P$, the
contribution of these $n$ agents to entanglement entropy remains
$(n\ln n)/(N\ln N)$ also at time $(t+1)$.
\begin{figure*}
  \begin{center}
    \includegraphics[width=1\linewidth]{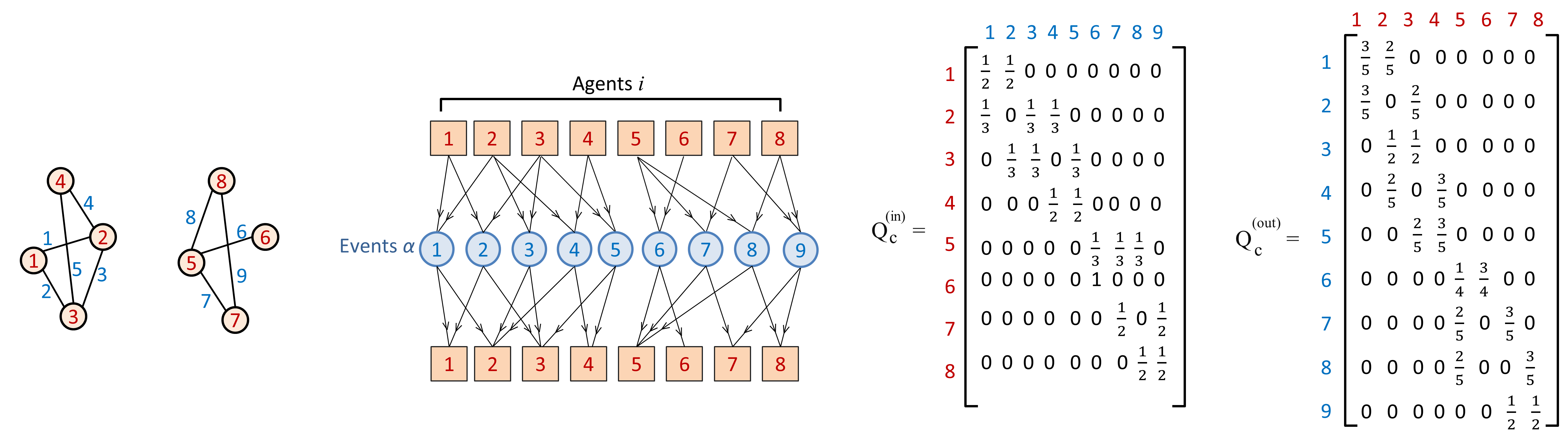}
    \caption{Constructing the $Q^{\text{(in)}}_{\text c}$ and the
      $Q^{\text{(out)}}_{\text c}$ for the working example, containing
      the agents' actual contact network at $t=0$ in Fig. 1(a) of the
      main text. \label{fine_structure}}
\end{center}
\end{figure*}

The method is trivially extended when there are multiple events taking
place at time $t$, as shown for an example in Fig. \ref{fragment}(b)
for 8 agents, which, along with the corresponding $P$ in
Fig. \ref{pfmatrix}, is then copied in Fig.~1(c) in the main text of
the paper.

\subsection{Entanglement entropy for the fine structures}

Calculation of entanglement entropy involving the fine
structures in the agents' contact networks at integer times can also
be performed by constructing the corresponding $Q^{\text{(in)}}$,
$Q^{\text{(out)}}$ and $P=Q^{\text{(in)}}\cdot Q^{\text{(out)}}$
matrices --- along very similar lines as in \ref{sec_a1}. The procedure is
in fact best illustrated by taking another working example, which we
do in Fig. \ref{fine_structure}: note that the fine structures
correspond to those shown in Fig. 1(a) at $t=0$ (in the main text of
the paper). The key difference between the two working examples in
Fig.~\ref{fragment} and in Fig.~\ref{fine_structure} is that in the
latter the fine structures of the agents' contact network at time $t$
is possible to capture only by binary events (i.e., each event
involves two agents, corresponding to the link between
them). Nevertheless, we follow the same convention for the denoting
the agents, the events, and the weights of the agent-to-event and
event-to-agent links as we did in \ref{sec_a1}.

First, the topology, as seen in Fig. \ref{fine_structure}, is again
time-reversal invariant, which allows us to write, analogous to
Eq.~(\ref{eA1}),
\begin{eqnarray}
  \frac{w_{i\alpha}}{w'_{\alpha i}}=\frac{w_{j\alpha}}{w'_{\alpha
  j}}=c_\alpha,
  \label{eB1}
\end{eqnarray}
where $i$ and $j$ are the two agents participating in (binary) event
$\alpha$, for some (yet unknown) $c_\alpha$. [That said, the topology
is no longer invariant under an exchange operation w.r.t. the agent
identities $i\leftrightarrow j$ as was the case in
Fig. \ref{fragment}(a)]. Next, we assign $w_{i\alpha}$ to be the
probability for a random walker to start at agent $i$ prior to time
$t$ and end at event $\alpha$ at time $t$. Using the principle that,
starting at agent $i$, the random walker chooses any of the connecting
events with equal probability, we obtain $Q^{\text{(in)}}_{\text c}$,
whose $i\alpha$-th element equals $w_{i\alpha}$
(Fig. \ref{fine_structure}). Finally, $Q^{\text{(out)}}_{\text c}$,
whose $\alpha j$-th element equals $w'_{\alpha j}$, and obeys
Eq. (\ref{eB1}), is obtained by transposing
$Q^{\text{(in)}}_{\text c}$ and subsequently row-normalizing it. The
matrix $Q^{\text{(out)}}_{\text c}$, obtained by means of this
procedure, is also shown in Fig. \ref{fine_structure}.
\begin{figure}[h]
  \begin{center}
    \includegraphics[width=0.8\linewidth]{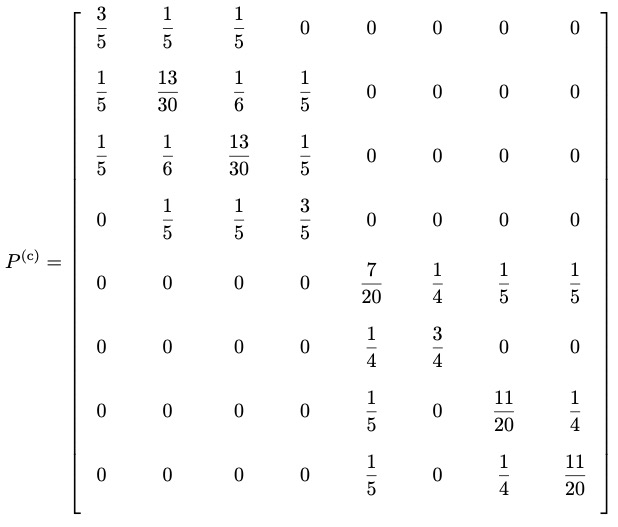}
  \caption{The $8\times8$ $P^{\text{(c)}}$-matrix constructed as $P^{\text{(c)}}=Q^{\text{(in)}}_{\text c}\cdot Q^{\text{(out)}}_{\text c}$ from Fig.~\ref{fine_structure}.\label{pmatrix}}
\end{center}
\end{figure}

The corresponding $P$ matrix, obtained as
$P^{\text{(c)}}=Q^{\text{(in)}}_{\text c}\cdot Q^{\text{(out)}}_{\text
  c}$ is shown in Fig. \ref{pmatrix}. It can be shown, using the above
procedure to determine $Q^{\text{(in)}}_c$ and $Q^{\text{(out)}}$,
that
\begin{eqnarray}
 \hspace{-8mm}P^{\text{(c)}}_{ij}=\frac{1}{{\mathcal K}_i+{\mathcal K}_j}\,\,\mbox{for $i\neq j$ and}\,\,
  P^{\text{(c)}}_{ii}=1-\sum_{j\neq i}P^{\text{(c)}}_{ij},
  \label{eB2}
\end{eqnarray}
where ${\mathcal K}_i$ is the degree of agent $i$ in the contact
network at time $t$. Clearly, $P^{\text{(c)}}_{ij}=0$ if agents $i$
and $j$ are not in contact at time $t$. Note here that
$P^{\text{(c)}}$, just as in $P$, is row- as well as column-normalized
(here too, the normalizations stem from conservation of random walkers
--- any random walker starting from any agent prior to time $t$ must
end up at some agent subsequent to time $t$). Once $P^{\text{(c)}}$ is
calculated in this way, the corresponding entanglement entropy
$S_{\text c}$ can be calculated following Eq.~(2) in the main text.

A discussion on the differences between the $P$ (Fig.~\ref{pfmatrix})
and the $P^{\text{(c)}}$ (Fig. \ref{pmatrix}) matrices is in
order. The difference between the two stems from the fact that the
agents within the two events in Fig. \ref{fragment}(b) are maximally
entangled, which is not the case for the agents in
Fig. \ref{fine_structure}. Mathematically, the difference lies in the
fact that $P^2=P$ in contrast to
$[P^{\text{(c)}}]^2\neq P^{\text{(c)}}$.

In a more precise formulation, we note that while constructing
$P^{\text{(c)}}$ by means of combining the probabilities of random
walkers to start at agent $i$ prior to time $t$ and end at event
$\alpha$ at time $t$, or start at event $\alpha$ at time $t$ and end
at agent $i$ subsequent to time $t$, we have allowed the random walker
to take one combined agent-to-agent hop across time $t$. From this,
one would expect that if a large number of such sequential
agent-to-agent random walker hops would be allowed while the agents'
contact network remains the same over time [i.e., the same agent
contact network holds at times $t, (t+1),\ldots,(t+k)$ for large $k$]
then the random walker starting prior to time $t$ at agent $i$ to
reach all agents within the corresponding connected component
subsequent to time $(t+k)$ with equal probability. In fact, this is
indeed the case, since
\begin{eqnarray}
  \lim_{k\rightarrow\infty}[P^{\text{(c)}}]^k = P.
  \label{eB3}
\end{eqnarray}
This limiting case is made possible by the condition that
$P^{\text{(c)}}$ matrices are row- as well as column-normalized (which
in essence is the detailed balance condition in statistical physics,
effected by the conservation of random walkers as stated
above \cite{VANKAMPEN2007193}).

Further reflection reveals that the subtleties regarding the
differences between $P$ and $P^{\text{(c)}}$ arise due to the
discreteness of the time snapshots. In the main text of the paper we
noted that real-world temporal network data are often sampled at some
fixed interval $\tau_s$, and also that the random walker hops are
coupled to $\tau_s$. Given that in theory, $\tau_s$ can be taken to be
infinitesimally small (in comparison to the time-scales of change in
the topology of the agents' contact network), and that in real world,
every interaction lasts for a finite amount of time, the limit
(\ref{eB3}) ensures that $P$ is the correct descriptor for measuring
entanglement.

\subsection{Contribution of a specific agent to entanglement}

The contribution $S_i$ of agent $i$ to the entanglement entropy is the
difference between the entanglement entropy when string $i$ is
embedded in the temporal network, and that when string $i$ is removed
from of the temporal network. Note that two entanglements of agent $i$
at two different shapshots, with agents $j$ and $k$ respectively,
makes agent $i$ an `in-between temporal stop' for agents $j$ and
$k$. With this in mind, $S_i$ can be seen as {\it a measure of
  betweenness of agent $i$ in a temporal network\/}.

The quantity $S_i$ can be calculated as follows. For every time
snapshot we construct the $P$ matrices, and the corresponding
$P^{(-i)}$ matrices by replacing all the $i$-th row and column
elements of the $P$ matrices by zeros, except the diagonal element
$P_{ii}$. We then row normalize both of them separately to obtain the
$P$ and the $P^{(-i)}$ matrices, and construct the product matrices
$\wp{(t,\Delta t)}\equiv P{(t)} P{(t+1)} \ldots P{(t+\Delta t)}$ and
$\wp^{(-i)}(t,t+\Delta t)\equiv P^{(-i)}(t) P^{(-i)}(t+1) \ldots
P^{(-i)}(t+\Delta t)$. This is followed by correspondingly calculating
$S(t,\Delta t)$ and $S^{(-i)}(t,\Delta t)$ using Eq.~2). This yields
us
\begin{eqnarray}
S_i(t,\Delta t)=S(t,\Delta t)-S^{(-i)}(t,\Delta t).
\label{eC1}
\end{eqnarray}

\section{Entanglement entropy as a function of $\Delta t$\label{sec_a2}}

In the main text, especially in Fig.~3, we have analyzed the evolution
of entanglement entropy. In this section, we investigate the highest
rate at which entropy can increase with the time interval size
$\Delta t$, which we in the end extrapolate to the special case of a
perfect tree --- i.e., a system we argue has the maximum possible
entropy growth with time.

Although in real-world networks, event sizes vary considerably, for
the sake of an illustration, given a certain initialization time $t$,
let us consider the case where the event sizes are constant ($=n$) for
$\Delta t=1$. Following Fig.~1, all rows of $\wp$ will have exactly
$n$ entries containing $\frac{1}{n}$, and $0$ elsewhere, leading to:
\begin{eqnarray*}
  S_{\text{tree}} (t, \Delta t = 1) &=& -\frac{1}{N \ln N} \cdot N \cdot n \cdot \left(\frac{1}{n} \ln \frac{1}{n}\right) \\
                                    &=& \frac{\ln n}{\ln N}.
\end{eqnarray*}

Extending this to higher values of $\Delta t$ requires more
formality. To this end, we consider an event ${\cal E}$ of size
$n_{\cal E}$ at some time $t'=t+\Delta t$ where $\Delta t>1$, and
conceptually imagine that agents from different `groups', labeled by
colors such as red, blue etc. participate. We define a group
${\cal G}$ as follows: in between times $t$ and $t'$, $\forall i$
belonging to one group, and $\forall j$ belonging to another, there is
no event where both $i$ and $j$ are temporally connected to. Prior to
time $t'$, let us denote the number of red agents (i.e., belonging to
the red group ${\cal G}_R$) by ${\cal N}_R$, the number of blue agents
(i.e., belonging to the blue group ${\cal G}_B$) by ${\cal N}_B$ and
so on. Similarly, the number of red agents participating in event
${\cal E}$ at time $t'$ is given by $n_R$ (forming the corresponding
subgroup $g_R$), the number of blue agents participating in event
${\cal E}$ at time $t'$ is given by $n_B$ (forming the corresponding
subgroup $g_B$), and so on. Clearly,
\begin{eqnarray}
n_R+n_B+\ldots = n.
\label{eD1}
\end{eqnarray}
We also label the elements of $\wp(t,\Delta t-1)$ belonging to a given
colored group by indices of the same color; e.g., we reindex the red
agents by red indices $1_R,2_R,\ldots,{\cal N}_R$, of which the agents
$1_R,2_R,\ldots,n_R$ participate in the event. Then the elements of
the $\wp(t,\Delta t-1)$ matrix corresponding to the red group of
agents can be extracted to form an ${\cal N}_R\times{\cal N}_R$
matrix. Let us denote these elements by the notation $p_{ij}$, with
$1_R\le (i,j)\le{\cal N}_R$. Then the following relations will hold:
\begin{eqnarray*}
\sum_{i=1_R}^{{\cal N}_R}\,p_{ij}(t-1)=1,\,\,i\in{\cal G}_R, j\in g_R;\\
  \sum_{i=1_B}^{{\cal
  N}_B}\,p_{ij}(t-1)=\ldots=1,\,\,i\in {\cal G}_B, j\in g_B;\,\,\ldots
\label{eD2}
\end{eqnarray*}
These relations simply follow from the fact that $\wp(t,\Delta t-1)$
is column-normalized, having noted that $\wp_{ij}(t,\Delta t-1)=0$
when agents $i$ and $j$ belong to two different(ly colored) groups.

\begin{widetext}
 Subsequent to the event at time $t$, for an agent $j\in{\cal E}$ the
elements of $P(t)$ will be given by
\begin{eqnarray}
  p_{ij}(t)=\frac{1}{n_{\cal E}}
  \begin{cases}
  \displaystyle{\sum_{j'=1_R}^{n_R}}\,p_{ij'}(t-1)& i\in {\cal G}_R, j'\in g_R\\\\
  \displaystyle{\sum_{j'=1_B}^{n_B}}\,p_{ij'}(t-1)& i\in {\cal G}_B, j'\in g_B\\\\
  \ldots
\end{cases},
  \label{eD3}
\end{eqnarray}
and is independent of $j$ [i.e., $p_{ij}(t)\equiv p_i(t)$]. This
allows us to express the corresponding change in entropy due to the
event ${\cal E}$, by separating the agents in different(ly colored)
subgroups, as
\begin{eqnarray}
  \Delta S&=&\frac{1}{N\ln N}\left\{\left[n_{\cal E}\sum_{i=1_R}^{{\cal N}_R}\,p_i(t) \ln p_i(t)-\sum_{i=1_R}^{{\cal N}_R}\sum_{j=1_R}^{n_R}\,p_{ij}(t-1) \ln p_{ij}(t-1)\right]\right.\nonumber\\ &&\hspace{1.6cm}\left.+\left[n_{\cal E}\sum_{i=1_B}^{{\cal N}_B}\,p_i(t) \ln p_i(t)-\sum_{i=1_B}^{{\cal N}_B}\sum_{j=1_B}^{n_B}\,p_{ij}(t-1) \ln p_{ij}(t-1)\right]+\ldots\right\}.
  \label{eD4}
\end{eqnarray}
Using Eq. (\ref{eD3}) to replace $p_{ij}(t)$ in Eq. (\ref{eD4}), and thereafter having dropped the $(t-1)$ argument for the matrix elements, we have
\begin{eqnarray}
 \Delta S &=& \frac{n_{\cal E}\ln n_{\cal E}}{N\ln N} + \frac{1}{N\ln N} \left\{\underbrace{\left[\sum_{i=1_R}^{{\cal N}_R}\sum_{j=1_R}^{n_R}\, p_{ij}\ln p_{ij}-\sum_{i=1_R}^{{\cal N}_R}\left(\sum_{j=1_R}^{n_R}\, p_{ij}\right)\ln \left(\sum_{j=1_R}^{n_R}\, p_{ij}\right)\right]}_{C_R}\right.\nonumber\\&&\hspace{3.3cm}\left.+\underbrace{\left[\sum_{i=1_B}^{{\cal N}_B}\sum_{j=1_B}^{n_B}\, p_{ij}\ln p_{ij}-\sum_{i=1_B}^{{\cal N}_B}\left(\sum_{j=1_B}^{n_B}\, p_{ij}\right)\ln \left(\sum_{j=1_B}^{n_B}\, p_{ij}\right) \right]}_{C_B} +\ldots\right\}.
  \label{eD5}
\end{eqnarray}
\end{widetext}
Given that $p_{ij}\le 1$, it is easily argued from Eq. (\ref{eD5})
that $C_R$, $C_B$ etc. terms are $\le 0$: the only case when they can
be zero is when $n_R$, $n_B$ etc. either $0$ or $1$. This in turn
means that an event can maximally contribute to the entanglement
entropy only when, for every agent pairs $(i,j)$ participating in the
event, between $t$ and $t'$ there exists no event that both $i$ and
$j$ are temporally connected to.

Note however that this condition is not possible to maintain for an
arbitrary temporal depth of between $t$ and $t+\Delta t$. A toy case
where this is possible --- returning to the one we started this
section with --- is with fixed event size $n$ and also fixed event
frequency $n$ per snapshot, requiring the condition $N=n^k$ to be
satisfied for some integer $k$. We illustrate the procedure with an
example below. When increasing $\Delta t$ to $2$ in that toy example,
we add an extra time step. For the system that is a perfect tree in
the temporal domain, any agents that have participated in events at
$\Delta t=1$, do not participate in an event at $\Delta t=2$; instead,
all $n$ agents from any event in the first time step participate in
$n$ separate events. This ensures that the all rows of the $\wp$
matrix have exactly $n^2$ elements with entries $\frac{1}{n^2}$,
resulting in $S(t, \Delta t = 2) = 2\frac{\ln n}{\ln N}$. Continuing
this for any value of $\Delta t$, this can be generalized to:
\begin{eqnarray*}
  S_{\text{tree}} (t, \Delta t)&=& -\frac{1}{N \ln N} \cdot N \cdot n^{\Delta t} \cdot \left(\frac{1}{n^{\Delta t}} \ln \frac{1}{n^{\Delta t}}\right) \\
                               &=&\Delta t \frac{\ln n}{\ln N},
\end{eqnarray*}
leading to a linear relationship between $S$ and $\Delta t$ for a
perfect tree in the temporal domain.

From a point of view of analyzing spreading dynamics, the perfect tree
is the system with the highest spreading vulnerability. For a perfect
tree in the temporal domain, agents only meet agents that do not have
a (higher-order) connection in common in the past. For the application
of rumor spreading, for example, in a perfect tree, people that
\textit{know} the rumor will therefore only meet people that
\textit{do not know} the rumor -- obtaining the fastest possible
spread of the rumor.

\section{Real-world temporal network datasets used in the paper\label{sec_a3}}

In Fig.~1 of the main text, we use a network constructed by hand for
illustration purposes. In Figs.~2 and 3 of the main text, we use
temporal network data from real-world systems. In this section we
discuss the preprocessing and availability of these datasets.

\subsection{High school and Primary school}

In Fig.~2 of the main text, we use data on the interactions among
students of a high school and a primary school. Both datasets are
frequently used in network science papers. The high school data is
referred to as the `Thiers13' dataset \cite{HighSchool_1}, from which
we use two days in the analysis, and the primary school data is the
`LyonSchool' dataset \cite{PrimarySchool_1, PrimarySchool_2}, from
which we also use two days of data in the table in Fig.~3(b) of the
main text. Both are freely accessible at the Sociopatterns project
website (\url{http://www.sociopatterns.org/datasets/}). Agent
interactions in these datases are defined as close face-to-face
proximity of students and teachers using wearable sensors and
proximity-sensing infrastructure based on radio frequency
identification devices. In the table in Fig.~3(b) of the main text, we
also use a second high school dataset (also from the Sociopatterns
project) from 2011, referred to as `Thiers11', from which we use four
days in the analysis. The sampling time interval $\tau_s$ for these
datasets is 20 seconds.

\subsection{Other Sociopatterns datasets}

Data from the scientific conference, hospital and workplaces (2013)
and (2015) are also accessible via the Sociopatterns project and are
often referred to as the SFHH conference, LH10, InVs13 and InVs15
datasets. The scientific conference refers to the 2009 SFHH conference
in Nice, France (June 4-5, 2009) \cite{Cattuto2010,Stehle2011b}. The
workplace datasets InVs13 and InVs15 were experiments conducted in
French office buildings in 2013 and 2015
respectively \cite{Genois2018}. The hospital dataset contains the
temporal network of contacts among patients and health-care workers
(HCWs) and among HCWs in a hospital ward in Lyon, France, from Monday,
December 6, 2010 at 1:00 pm to Friday, December 10, 2010 at 2:00 pm
--- we use four separated days in our analysis from this time
interval. The study included 46 HCWs and 29
patients \cite{hospital}. The sampling time interval $\tau_s$ for these
datasets is 20 seconds.

\subsection{High-energy physics citations}

The citations network we analyze in the paper is from the e-print
arXiv server HEP-PH (high energy physics phenomenology) and covers all
the citations within a dataset of 34,546 papers with 421,578
edges\cite{Citations}. Because of the enormous amount of agents (i.e.,
papers) and (temporal) links, we only use a subset of this
dataset. The dataset comprises papers over approximately 10 years of
data. We only use papers between 11 March 1996 and 3 Aug 1998. These
dates are obtained by taking 10,000 interactions in approximately the
middle of the time series (to best relate to the temporal dynamics of
this system). This results in a much smaller number of 4,052 papers
(approximately 10\% of the total). If a paper $i$ cites paper $j$, we
treat this as an `interaction' between these two papers. The sampling
time interval $\tau_s$ for these datasets is in days.

\subsection{Baboons}

The baboons temporal network we analyzed is from an experiment on a
group of 20 Guinea baboons living in an enclosure of a Primate Center
in France, between June 13th 2019 and July 10th
2019 \cite{baboons}. Only 13 out of the 20 baboons wore proximity
sensors -- which is the data we use for this analysis. In our
analysis, we use the first five days of this dataset. The sampling
time interval $\tau_s$ for these datasets is 20 seconds.

\subsection{Sparrows}

The sparrows data set is from a study at the University of California,
Santa Cruz Arboretum \cite{Sparrows}. The sparrows arrive there in
October-November and depart for their breeding grounds in March-April
each year. The study spanned three non-breeding seasons: January-March
2010 (Season 1), October 2010-February 2011 (Season 2) and October
2011-April 2012 (Season 3). Each season, the authors captured birds
using baited traps and attached individually unique combinations of
color bands. In Season 2, they did not band any birds between
October-December 2010. Network edges are found based on co-memberships
between flocks (defined as a group of individuals found within a
single 5m radius) by identifying the color-banded individuals in each
flock. Data are available at $\tau_s=1$ day.

\subsection{Irvine social app}

This dataset is comprised of private messages sent on an online social
network at the University of California, Irvine. Users could search
the network for others and then initiate conversation based on profile
information. An edge $(i, j, t)$ means that user $i$ sent a private
message to user $j$ at time $t$ \cite{Irvine}. The data as a whole
spans of approximately 195 days. As with the aforementioned Citations
dataset, for this analysis, we focused on only part of the dataset to
improve computational speed. In particular, we used the first 34 days,
resulting in a set of 25,000 interactions and 1136 unique agents
(users). The time is aggregated to $\tau_s=6$ hrs.

\subsection{Art fair}

The art fair data is gathered during `Smart Distance Lab: The Art
Fair' between August 28 and 30, 2020 in Amsterdam, the
Netherlands \cite{Tanis2020}. It consists of eight experiments, where
various conditions such as walking direction, face masks and proximity
alerting systems were varied. We use the data from the wearable
sensors, that defined an interaction by two people coming in the
proximity of less than 1.5 m. The data set is available on Figshare
and in a MySQL database. The sampling time interval $\tau_s$ for this
dataset is one second.

\section{Spreading processes on top of temporal networks\label{sec_a4}}

In order to reveal the relationship between the entanglement entropy
and the dynamical processes playing out on top of the network, we
simulate three models describing three different (stochastic)
processes on the real-world systems in Fig.~3 and Tab.~1 of the main
text. The results of the simulations, reported in the main text, have
been obtained by performing ensembles of 25 re-initialized model runs
(per initialization time $t$) over 80\% of the full time window per
system as starting points $t$, using $\Delta t = $ 12 time steps for
the correlations in Fig.~3b in the main text, and 60 time steps for
the visualizations in panels (c) and (d). Only ensemble averages are
reported in the main text. For most datasets, this procedure is
repeated over multiple subsets. For example, the high school data
Thiers13 contains five days of data. Those days are individually run
using this ensemble procedure, and the average across those days is
shown in the table in Fig.~3(b) of the main text. Below, we describe
the stochastic models. For all models, the number of agents is denoted
by $N$.

\subsection{Majority-vote}

For the first model we simulate the propagation of opinions on a
dilemma, which is associated to spreading of information, such as
(fake) news, memes or political opinions. The opinion of agent $i$ at
time $t$ can have values $0$ or $1$. We denote the overall fraction of
agents having opinion $0$ with $f_0$, and analogously we define
$f_1$. At initialization (starting time $t$, interval size
$\Delta t = 0$), we start with half of the population having opinion
$0$, and the other half opinion $1$. The agent interactions are
specified as they are defined by the (real-world) temporal network's
topology. At each event, all agent's change their opinion to the
opinion of the majority of the participants. For example, if in an
event with three agents, the opinions are $001$, then, the third agent
will change its opinion to $0$. If the opinions are tied, all
participants will swap to a single opinion, chosen at random. If there
is only one single agent in an event, its opinion will remain the
same. No noise is introduced. We define a single state variable that
is dependent on the starting time of the model ($t$) and the time
interval ($\Delta t$) that is used to progress it from $t$ onward:
\begin{eqnarray*}
V_{\text{maj}}(t, \Delta t) &=& 2\cdot [\text{max}(f_0(t, \Delta t), f_1(t, \Delta t))-0.5].
\end{eqnarray*}
The subtraction of 0.5 and subsequent multiplication with 2 make sure
that $V_{\text{maj}}\in[0, 1]$, which is a standardized form for easy
comparison to the entanglement entropy.
\begin{figure*}
  \includegraphics[width=1\linewidth]{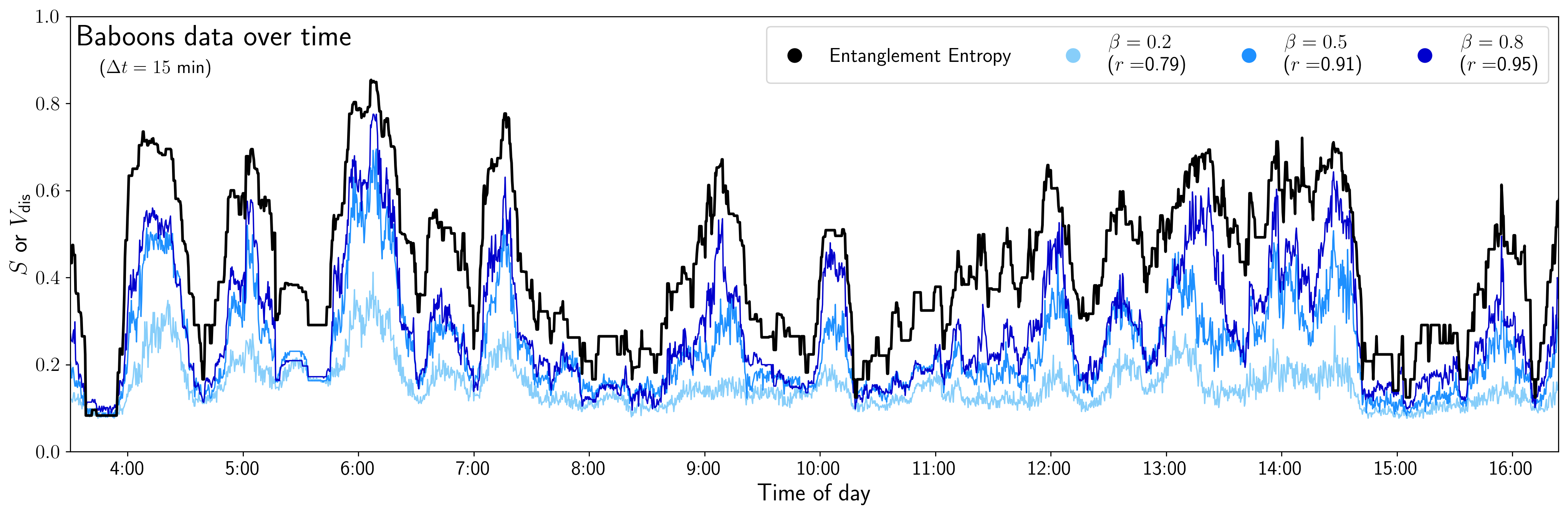}
  \caption{Evolution of entanglement entropy $S$, and the state
    variable $V_{\text{dis}}$, for three values of $\beta$, using
    $\Delta t = 15$ minutes, for day 2 in the Baboons
    dataset. Pearson's correlation values ($r$) are also noted in the
    figure (which are the values only for this particular day). Note
    that Fig.~3(b) of the main text uses five days of the Baboons
    dataset, leading to a slightly smaller Pearson's correlation value
    for $\beta=0.8$, namely 0.93.}
\end{figure*}

\subsection{Transport delay}

The second model concerns the propagation of a continuous variable
through interactions (events), all event participants will attain the
highest value among them. This relates to various real-world
phenomena, such as transportation delays (hence the title of this
model), in which transport assets (such as a physical train and a
driver to drive it; assets may need to come from different physical
locations \cite{Dekker2021plos}, and have to wait for the
maximally-delayed one --- assigning all assets in an event with a
newly generated delay that equals the highest of all the delays in
that event. (Another example is the spread of dominant genes in
phylogenetics.) At the start of the model, all agents are given a
random number between 0 and 1, we refer to as the `delay' $d_i$ of
agent $i$, which propagates as per the above rule. Over time, the
average delay will increase up to the maximum existing delay. We
define a state variable, again dependent on $t$ and $\Delta t$:
\begin{eqnarray*}
  V_{\text{del}}(t, \Delta t) &=& 2\cdot \left[\frac{1}{N}\sum_{i=1}^N d_{i=1}(t, \Delta t)-0.5\right],
\end{eqnarray*}
which has the same standardization as the state variable of the
majority-vote system, to bring $V_{\text{del}}$ between $0$ and $1$.

\subsection{Infectious disease}

While numerous infectious disease models for networks exist, we chose
the simple susceptible-infected (SI) model. Each of the model's agents
$i$ can be in either one of two states $q_i$: susceptible ($q_i = 0$)
or infected ($q_i = 1$). There is no exposed or recovered state, as
commonly used in epidemiology. Each model run is initialized with 10\%
(randomly chosen) agents being infected. Upon interaction in an event,
if one of the participants is infected, there is a probability $\beta$
of other (susceptible) participants becoming infected as well. We
define a state variable dependent on $t$ and $\Delta t$, which is the
average status $q$ over all agents:
\begin{eqnarray*}
V_{\text{dis}}(t, \Delta t) &=& \frac{1}{N}\sum_{i=1}^N q_i(t, \Delta t),
\end{eqnarray*}
which is a number between $0$ (no infected) and $1$ (all agents infected).

Clearly $\beta$ is a parameter for the spreading of the pathogen. For
the main text of the paper, we have chosen $\beta = 0.8$. For the
Baboons dataset, below we present the data for two additional values
of $\beta$. While the Pearson's correlation between $S$ and
$V_{\text{dis}}$ decreases with decreasing $\beta$, the correlation
still remains substantial even for $\beta$ as low as $0.2$.



\end{document}